\documentclass[twocolumn,aps,amsmath,amssymb,pra]{revtex4}
\usepackage{bm}

\newcommand{\B}{\mbox{\tiny B}}
\newcommand{\D}{\mbox{\tiny D}}
\newcommand{\ti}{\tilde}
\newcommand{\nl}{\nonumber \\}

\newcommand{\Sec}[1]{Sec.\,\ref{#1}}
\newcommand{\App}[1]{Appendix\,\ref{#1}}
\newcommand{\be}{\begin{equation}}
\newcommand{\ee}{\end{equation}}
\newcommand{\bea}{\begin{eqnarray}}
\newcommand{\eea}{\end{eqnarray}}
\newcommand{\bsube}{\begin{subequations}}
\newcommand{\esube}{\end{subequations}}
\newcommand{\Eq}[1]{Eq.\,(\ref{#1})}
\newcommand{\Eqs}[1]{Eqs.\,(\ref{#1})}

\newcommand{\dg}{\dagger}
\newcommand{\la}{\langle}
\newcommand{\ra}{\rangle}

\newcommand{\ind}{{\sf n}}
\newcommand{\bn}{\bar n}
\newcommand{\bfA}{\bm A}
\newcommand{\bfB}{\bm B}
\newcommand{\bfalp}{\bm\alpha}
\newcommand{\bfLam}{\bm\Lambda}
\newcommand{\bfPi}{\bm\Pi}
\newcommand{\rvec}{\mbox{\boldmath ${\rho}$}}
\newcommand{\Uvec}{\mbox{\boldmath ${\cal U}$}}
\newcommand{\Gvec}{\mbox{\boldmath ${\cal G}$}}

\newcommand{\rhonswap}{\rho_{\sf n}^{{ }_{\{\leftrightarrows\}}}}
\newcommand{\rhonup}{\rho_{\sf n}^{{ }_{\{+\}}}}
\newcommand{\rhondown}{\rho_{\sf n}^{{ }_{\{-\}}}}

\newcommand{\rhoNup}{\rho_{\sf N}^{{ }_{\{+\}}}}

\begin{document}

\title{Dynamics of quantum dissipation systems
  interacting with bosonic canonical bath:
      Hierarchical equations of motion approach
}

\author{Rui-Xue Xu and YiJing Yan}
\affiliation{Hefei National Laboratory for Physical Sciences at
    the Microscale,
  University of Science and Technology of China, Hefei, China, and
  \\
  Department of Chemistry,
  Hong Kong University of Science and Technology, Kowloon, Hong Kong}

\date{Submitted on 24 October 2006, Revised on 11 December 2006}

\begin{abstract}
  A nonperturbative theory is developed, aiming
 at an exact and efficient evaluation of a general
 quantum system interacting with arbitrary bath
 environment at any temperature and in the presence
 of arbitrary time-dependent external fields.
 An exact hierarchical equations of motion formalism
 is constructed on the basis of calculus-on-path-integral algorithm,
 via the auxiliary influence
 generating functionals related to
 the interaction bath correlation functions
 in a parametrization expansion form.
 The corresponding continued-fraction Green's functions
 formalism for quantum dissipation is also presented.
 Proposed further is the principle of residue correction,
 not just for truncating the infinite hierarchy,
 but also for incorporating the small residue dissipation
 that may arise from the practical difference
 between the true and the parametrized bath correlation functions.
  The final residue-corrected hierarchical equations
 of  motion can therefore be used practically
 for the evaluation of arbitrary dissipative quantum systems.
\end{abstract}

\maketitle
\section{Introduction}
    As a fundamental topic in quantum statistical mechanics,
 the development of quantum dissipation theory
 has involved scientists from diversified fields of
 research over decades \cite{Fey63118,Red651,Kub85,%
  Wei99,Kle06,Ple98101,Dit98,Tho0115,Bre02,%
 Gra88115,Cal83374,Leg871,Yan982721,%
  Yan002068,Xu029196,Yan05187,Xu05041103,Mak952430,%
 Tan89101,Tan914131,Ish053131,Tan06082001,%
 Sha045053,Yan04216,Sha06187,Sto994983,Sto02170407,%
 Str04052115,Yu04062107,Han05026105}.
 The key quantity in quantum dissipation theory
 is the reduced density operator,
 $\rho(t) \equiv {\rm tr}_{\B}\rho_{\rm T}(t)$, i.e.,
 the partial trace of the total density operator over
 the bath space of practically infinite degrees of freedom.
 The standard approach to the exact $\rho(t)$
 is the influence functional path integral formalism \cite{Fey63118}.
 However, from both operational and numerical points of view,
 this formally exact formalism is much limited
 in comparison with its differential or master equation counterpart.

   The attempt to the construction of quantum master equation,
 via the derivative on the exact path integral formalism, was
 first carried out by Caldeira and Leggett,
 for the quantum dissipation
 in the high temperature Markovian limit \cite{Cal83374}.
 The calculus--on--path--integral (COPI) method
 has also been used in driven bistable systems,
 together with the so-called non-interacting
 blip approximation or its variations to treat the quantum
 path correlations
 in the reduced dynamics \cite{Leg871,Wei99,Tho0115}.
%
 The first exact quantum master equation
 via the COPI method is constructed for the damped harmonic
 oscillator systems \cite{Hu922843,Kar97153}.
  Tanimura and coworkers have recently constructed
 a set of hierarchically coupled quantum Fokker-Planck equations
 for a general Drude dissipation system \cite{Ish053131},
 without any longer the high-temperature approximation
 exploited before \cite{Tan89101,Tan914131}.
 The deterministic description of
 an exact differential equations of motion (EOM) formalism can also be
 constructed via stochastic descriptions of quantum dissipation
 \cite{Tan06082001,Sha045053,Yan04216,%
 Sha06187,Sto994983,Sto02170407,Str04052115,Yu04062107}.
 However, it has by far only been carried out
 with a certain single-mode dissipation model system.
 The general theory of quantum dissipation
 in terms of deterministic EOM remains a great challenge.

    The aim of this work is right at this fundamental issue
 of quantum statistical mechanics. It is to develop an
 exact and efficient EOM formalism
 for a general quantum system, interacting with arbitrary bath
 environment at any temperature, and arbitrary
 time-dependent external fields. The only assumption
 involved is the same as in the path integral
 formalism that the interaction bath satisfies the
 Gaussian statistics \cite{Fey63118,Wei99,Xu05041103}.
 The theoretical construction will be made mainly based on
 the COPI algorithm developed in Ref.\ \onlinecite{Xu05041103}.
 A set of so-called influence generating functionals
 will be shown to be crucial in formulating the
 exact hierarchical EOM.
 The present work focuses on the canonical bath ensemble case,
 but can be easily extended to the grand
  canonical case \cite{Jin2beJCP}.

 The remainder of this paper is organized as follows.
 In \Sec{thpathint}, after the description of the general form of
 system--bath coupling Hamiltonian,
 we reviews the path integral
influence functional formalism \cite{Xu05041103},
together with its formal relation to the time-local
dissipation superoperator in quantum master equation.
%
 In \Sec{thdebye}, we illustrate with the Drude--Debye model
 the key technical issue on the construction
 of hierarchical EOM. It is to
 identify a complete set of auxiliary
 Influence Generating Functionals via the
 COPI algorithm \cite{Xu05041103},
 which will be termed as the IGF--COPI for short hereafter.
 Section \ref{theom} turns to general dissipative quantum systems.
 It involves a parametrization of general
 interaction bath correlation functions that satisfy
 the fluctuation--dissipation theorem; see \App{thapp_para} for
 details. Upon identifying the complete set of auxiliary IGF--COPI
 construction for the parametrized forms of
 correlation functions, exact hierarchical EOM are followed
 immediately.
 The equivalent formalism in terms of hierarchical and continued-fraction
 Green's functions and memory kernels are given in \App{thapp_frac}.
 In \Sec{thresidue}, we further establish
 the so-called principle of residue correction.
 The principle itself is rooted at
 the formally exact relation established  in \Sec{thpathint},
 between the Feynman-Vernon influence functional
 and the time-local dissipation superoperator.
 But it is now exploited in certain perturbative or nonperturbative
 approximate manner to construct an efficient
 hierarchy truncation method.
 It is also used to incorporate the residue contribution, due to
 the difference between the exact and the parametrized bath,
 into the final theory.
 Finally, \Sec{thsum} concludes this paper.

\section{Quantum dissipation in terms of influence functionals}
\label{thpathint}

\subsection{Multi-mode system--bath coupling Hamiltonian}
\label{thpathintA}

 Consider a general quantum system embedded in a bath which
 is assumed as a canonical ensemble in this work.
 The total system--plus--bath composite Hamiltonian
 can be written in general as
 \be \label{HT0}
    H_{\rm T} = H(t) - \sum_a Q_a \hat F_a + h_{\B}.
 \ee
  Here, $H(t)$ and $h_{\B}$ denote the uncorrelated
  system and bath Hamiltonians, respectively.
 The former may be subject to a time-dependent coherent field drive.
 The second term in the right-hand-side (rhs)
 of \Eq{HT0} denotes the multi-mode system--bath coupling.
 It can be generally expressed in the multiple dissipative mode
 decomposition form,  in which $\{Q_a\}$ and $\{\hat F_a\}$
 are the system and bath operators, respectively,
 and assumed to be Hermite in this work.

  For the later use, introduce here the
 Liouvillian ${\cal L}$ and the dissipative coupling
 mode ${\cal Q}_a$ in the reduced system subspace
 via their actions on an arbitrary operator as
\be \label{calLQ}
 {\cal L}\hat O\equiv[H(t),\hat O], \ \
 {\cal Q}_a \hat O \equiv [Q_a,\hat O] .
\ee
Throughout this work, we set $\hbar \equiv 1$ and the inverse
temperature $\beta \equiv 1/(k_{\B}T)$, and denote
also $\partial_t\equiv \partial/\partial t$.

  The stochastic canonical bath ensemble
 average of an operator $\hat O$ is denoted as
\be \label{aveB}
 \la\hat O\,\ra_{\B}\equiv{\rm tr}_{\B}(\hat O \rho^{\rm eq}_{\B})
  = {\rm tr}_{\B}(\hat O
   e^{-\beta h_{\B}})/\text{tr}_{\B}e^{-\beta h_{\B}}.
\ee
 In the $h_{\B}$--interaction picture,
 \be \label{FtHeis}
   \hat F_a(t) \equiv e^{ih_{\B}t} \hat F_a e^{-ih_{\B}t},
 \ee
for each bath interaction operator in \Eq{HT0},
is assumed a Gaussian stochastic process.
 This is exactly the case
 when the bath consists of a collection of harmonic
 oscillators and each $\hat F_a$  is a linear combination
 of the coordinates and momenta of the harmonic bath oscillators.
 The Gaussian stochastic process is also related to the
 central limit theorem in statistics.
 As the quantity $Q_{a}\la \hat F_{a}\ra_{\B}$,
 if it is not zero,
 can be included in the system Hamiltonian in \Eq{HT0},
 we can set $\la \hat F_{a} \ra_{\B} =0$
 without loss of generality.
 The effects of Gaussian stochastic bath operators
 are then completely described by their correlation functions,
 \be \label{FFCorr}
  C_{ab}(t-\tau) = \la \hat F_{a}(t) \hat F_{b}(\tau) \ra_{\B}.
 \ee
 They satisfy the symmetry and detailed-balance relations
 of the canonical bath \cite{Yan05187,Wei99},
 \be \label{CtsymFDT}
   C^{\ast}_{ab}(t) = C_{ba}(-t) = C_{ab}(t-i\beta).
 \ee

  We shall be interested in a differential EOM formalism, aiming at an
 efficient and exact evaluation of the
 reduced dynamics, with arbitrary multi-mode non-Markovian dissipation
 and time-dependent external fields on the system.
  The key quantity
 of interest is the reduced
 density operator $\rho(t)$, defined together with its
 associating propagator ${\cal U}(t,t_0)$ as
 \be \label{rhot_def}
   \rho(t)  \equiv  {\rm tr}_{\B}\rho_{\rm T}(t)
  \equiv {\cal U}(t,t_0)\rho(t_0) .
 \ee
 The desired EOM will be constructed (cf.\ \Sec{theom}) via
 the IGF-COPI approach \cite{Xu05041103}, starting from
 the exact path integral expression that involves
  only the initial factorization ansatz
  $\rho_{\rm T}(t_0)=\rho(t_0)\rho^{\rm eq}_{\B}$.
 Note that when the initial time is set to $t_0\rightarrow -\infty$,
 this ansatz becomes exact \cite{Wei99,Yan05187}.

\subsection{Reduced dynamics versus influence functionals}
\label{thpathintB}

  This subsection summarizes the
 path integral formalism of quantum dissipation
 for the multi-mode system-bath interaction \cite{Xu05041103}.
 Exploited explicitly is only the Gaussian statistical
 property, the essence of linear harmonic bath coupling
 with arbitrary system operators $\{Q_a\}$.
 Unlike the EOM formalism that can be expressed in the operator level,
 the path integral expression goes with a representation.
 Let $\{|\alpha\ra\}$ be a basis set in the system subspace.
 In the $\alpha$-representation, \Eq{rhot_def} reads
 [setting ${\bm \alpha}\equiv(\alpha,\alpha')$ for abbreviation]
\be \label{rhotPI_def}
 \rho({\bm \alpha},t) \equiv \rho(\alpha,\alpha'\!,t)
 = \!\int\!d{\bm \alpha}_0\, {\cal U}({\bm \alpha},t;{\bm \alpha}_0,t_0)
 \rho({\bm \alpha}_0,t_0).
\ee
The reduced Liouville-space propagator reads
in terms of path integral as \cite{Fey63118}
 \bea \label{calGPI}
   {\cal U}(\bfalp,t;\bfalp_0,t_0)
 = \int_{\bfalp_0[t_0]}^{\bfalp[t]}   \!\!  {\cal D}{\bfalp} \,
     e^{iS[\alpha]} {\cal F}[\bfalp] e^{-iS[\alpha']}.
 \eea
The effect of system--bath interaction on the reduced system dynamics
is described by the Feynman--Vernon influence
functional ${\cal F}$, which will be elaborated soon.
In \Eq{calGPI}, $S[\alpha]$ is the classical action functional
of the reduced system,
evaluated along the path $\alpha(\tau)$, with the constraints of two
ending points $\alpha(t_0)=\alpha_0$ and
$\alpha(t)=\alpha$ being {\it fixed}.
In the absence of bath interaction (${\cal F}=1$),
the dynamics would be completely coherent; i.e.,
$\partial_t {\cal U} = -i{\cal L}\,{\cal U}$; if ${\cal F}=1$.

  Consider now the key quantity, the bath interaction
 induced influence functional ${\cal F}$.
 Traditionally, its expression is derived by adopting a single-mode
 system-bath interaction model \cite{Fey63118,Wei99},
 in which the bath $h_{\B}$
 is assumed to consist of a set of uncoupled harmonic oscillators
 $\{q_j\}$ and the system-bath interaction assumes the form of
 $H'=-Q\hat F=-Q\sum_j c_j q_j$, rather than the multi-mode decomposition
 as the second term of \Eq{HT0}.

  In connection to the later development of EOM formalism,
 we denote ${\bm a}=(aa')$ for a pair of dissipation modes hereafter
 and introduce [cf.\ the Eqs.\ (5) and (6) of Ref.\ \onlinecite{Xu05041103}]
 \be  \label{ticalQa}
  \ti{\cal Q}_{\bm a}(t;\{\bfalp\})
  \equiv
  \ti Q_{aa'}(t;\{\alpha\}) - \ti Q^{\prime}_{aa'}(t;\{\alpha'\}),
 \ee
 where (noting that $\ti Q_{\bm a}\equiv \ti Q_{aa'}$ and
 $C_{\bm a}\equiv C_{aa'}$)
  \bsube \label{tilQFI}
 \bea
 \ti Q_{\bm a}(t;\{\alpha\})
 &\!\!\equiv&\!\!
  \int_{t_0}^{t}\!d\tau\,
  C_{\bm a}(t-\tau) Q_{a'}[\alpha(\tau)],
 \label{tilQa}\\
  \ti Q^{\prime}_{\bm a}(t;\{\alpha'\})
 &\!\!\equiv&\!\!
  \int_{t_0}^{t}\!d\tau\,
  C^{\ast}_{\bm a}(t-\tau) Q_{a'}[\alpha'(\tau)].
 \label{tilQprime}
 \eea
 \esube
 Denote also
 \be \label{calQa}
  {\cal Q}_a[\bfalp(t)] \equiv  Q_a[\alpha(t)]-Q_a[\alpha'(t)].
 \ee
 It is in fact the $Q_a$-commutator [cf.\ \Eq{calLQ} the second identity]
 in the path integral representation,
 as it depends only on the {\it fixed} ending points.
 The final expression of the
 influence functional reads \cite{Xu05041103}
 \bsube \label{FV_FPhi}
 \be \label{FV_FPhiA}
   {\cal F}[\bfalp] \equiv
  \exp\left\{-\int_{t_0}^t\!\!d\tau\,
  {\cal R}[\tau;\{\bfalp\}]\right\},
 \ee
 with
 \bea \label{calR_PI}
  {\cal R}[t;\{\bfalp\}] \equiv \sum_{\bm a} {\cal Q}_a[\bfalp(t)]\,
    \ti{\cal Q}_{\bm a}(t;\{\bfalp\}).
\eea
\esube
 The above relations will be used together
 with \Eqs{ticalQa}--(\ref{calQa}) in the
 following sections to develop the desired EOM.

\subsection{Influence functional versus dissipation superoperator}
\label{thpathintC}

 Note that in \Eq{FV_FPhi} the conventional
 influence phase functional is now expressed in terms  of
 its time integrand ${\cal R}$. The latter is in fact
 the time-local dissipation superoperator ${\cal R}(t)$
 in the path integral representation, as the
 time derivative of \Eq{calGPI} with \Eq{FV_FPhi} leads to
 \bsube \label{POP}
 \be \label{POP_U}
   \partial_t {\cal U} = -i{\cal L}\,{\cal U} -{\cal R}(t){\cal U},
 \ee
 or equivalently
 \be \label{POP_rho}
   \dot\rho = -i{\cal L}\rho - {\cal R}(t)\rho.
 \ee
 \esube
 We can therefore call ${\cal R}$ of \Eq{calR_PI}
 as the dissipation functional.
 As inferred from \Eq{calR_PI} with \Eqs{ticalQa} and (\ref{calQa}),
 it may be symbolically expressed
 in the operator level as
 \be \label{calR_def}
  {\cal R} \hat O = \sum_{\bm a}
  [Q_a,  \ti Q_{\bm a}\hat O - \hat O \ti Q_{\bm a}^{\dg}].
 \ee
 Here, $\ti Q_{\bm a}$ denotes the operator form of \Eq{tilQa}.
 However, the explicit operator-level expression for
 ${\cal R}$ (or $\ti Q_{\bm a}$) is generally not available.
 In the presence of a time-dependent external field,
 the only case being of analytical expression of ${\cal R}$
 (or ${\ti Q}_a$) is,  to our best knowledge,
 the driven Brownian oscillator system \cite{Yan05187}.

  The formal relations, \Eqs{FV_FPhi}--(\ref{calR_def}),
 also highlight where the difficulty is in the exact evaluation
 of quantum dissipation; all relate to the memory-containing
 $\ti Q$- or equivalent $\ti{\cal Q}$-functionals.
 In a certain sense, the EOM formalism to be presented in
 detail soon (cf.\ \Sec{thdebye} and \Sec{theom})
  is a mathematical construction that hierarchically
 resolves the ``history'' containing in the
 $\ti{\cal Q}$-functionals.
 This issue will become evident in the coming sections.
 Moreover, these formal relations may also be directly
 useful for the residue correction of the final formalism,
 due to either hierarchy truncation, or the small
 difference between the exact but complicated
 bath correlation functions and the
 parametrized ones (cf.\ \Sec{thresidue}).

\section{Hierarchical equations of motion: Correlated Debye dissipation}
\label{thdebye}
  Before presenting the formalism for arbitrary dissipative systems,
let us consider in this section
the simplest multi-mode dissipation case, the Drude-Debye model,
in which
\be\label{debyeC}
  C_{\bm a}(t>0) = C_{aa'}(t)=\eta_{\bm a}e^{-\gamma_{\bm a}t}.
\ee
The parameters $\gamma_{\bm a}\equiv \gamma_{aa'}=\gamma_{a'a}$
are real, while $\eta_{\bm a}\equiv \eta_{aa'}=\eta_{a'a}^\ast$
are complex, and they are the same as the Drude parameters
$\gamma^{\bm a}_{\D}$ and $\eta^{\bm a}_{\D}$
in the next section, where the general case is studied.
Despite this fact, this section provides with clarity
the basic ingredient of the IGF--COPI approach to
the desired hierarchical EOM.

 The hierarchy construction starts with the time derivative
on the propagator ${\cal U}$ [\Eq{calGPI}] of the primary interest.
The time derivative on the action functional parts contributes to
the coherent dynamics of $-i{\cal L}U$, and thus
can be included into the final EOM.
We shall show in the following that the hierarchy
generation stems from the time derivative
on the evolving influence functionals in each tier
of the construction.

  Consider first the time derivative on the influence
functional of primary interest [\Eq{FV_FPhi}].
\be \label{dotF_debye0}
  \partial_t {\cal F}[\bfalp]
 = -\Big[\sum_{\bm a}{\cal Q}_a[\bfalp(t)]
   \ti{\cal Q}_{\bm a}(t;\{\bfalp\})\Big]{\cal F}[\bfalp].
\ee
 We shall hereafter omit the explicit path integral variables dependence
whenever it does not cause confusion. As results,
we recast \Eq{dotF_debye0} as
\be \label{dotF_debye}
  \partial_t {\cal F}=
  -i\Big[\sum_{\bm a}{\cal Q}_a(-i\ti{\cal Q}_{\bm a})\Big]{\cal F}
 \equiv -i\sum_{\bm a}{\cal Q}_a {\cal F}_{\bm a}.
\ee
The last identity introduces the
{\it auxiliary influence functionals} (AIFs),
\be \label{F1_debye}
  {\cal F}_{\bm a} \equiv (-i\ti{\cal Q}_{\bm a}){\cal F}.
\ee
In contrast to ${\cal F}$ whose leading term is 1,
the first tier AIFs \{${\cal F}_{\bm a}$\}
are of the second-order in the system--bath coupling
as their leading terms.
The hierarchy to be constructed will go naturally with
increasing the order of system--bath coupling.

 Consider now the time derivative on the first tier AIFs [\Eq{F1_debye}].
\be \label{dotF1_debye}
  \partial_t{\cal F}_{\bm a} =
 -i(\partial_t\ti{\cal Q}_a){\cal F}
 -i\Big[\sum_{\bm b}{\cal Q}_b (-i\ti{\cal Q}_{\bm a})
   (-i\ti{\cal Q}_{\bm b})\Big]{\cal F} .
\ee
The second term, which arises from the derivative on the primary
${\cal F}$ [cf.\ \Eq{dotF_debye}],
introduces now a set of new AIFs,
\be \label{F2_debye}
  {\cal F}_{{\bm a}{\bm b}} \equiv
    (-i\ti{\cal Q}_{\bm a}) (-i\ti{\cal Q}_{\bm b}){\cal F}.
\ee
The leading terms in these second-tier AIFs
 are of the fourth-order system--bath coupling.

 The first term of \Eq{dotF1_debye} involves
the time derivative on $\ti{\cal Q}_a$ [\Eq{ticalQa} with \Eqs{tilQFI}].
From \Eq{tilQa}
\be \label{dottiQ_debye}
 \partial_t{\ti Q}_{\bm a}=C_{\bm a}(0)Q_{a'}[\alpha(t)]
  + \ti{\ti Q}_{\bm a}(t;\{\alpha\}),
\ee
with
\be \label{titiQ}
 \ti{\ti Q}_{\bm a}(t;\{\alpha\})
 \equiv \int_{t_0}^t\!d\tau\, \dot{C}_{\bm a}(t-\tau)Q_{a'}[\alpha(\tau)].
\ee
Apparently, $\ti{\ti Q}_{\bm a}$ leads
to \Eq{dotF1_debye} a non-hierarchy term in general,
unless a specific form of $C_{\bm a}(t)$ such as \Eq{debyeC}
is considered. In the present case of study,
$\ti{\ti Q}_{\bm a}=-\gamma_{\bm a}{\ti Q}_{\bm a}$,
leading to
\be \label{dotcalQ_debye}
  \partial_t{\ti{\cal Q}_a}=i{\cal C}_{\bm a}
   - \gamma_{\bm a}\ti{\cal Q}_a.
\ee
Here [noting that $C_{\bm a}(0)=\eta_{\bm a}$ from \Eq{debyeC}]
\be \label{calCPI_debye}
  {\cal C}_{\bm a}=-i\big\{\eta_{\bm a}Q_{a'}[\alpha(t)]-
      \eta_{\bm a}^{\ast}Q_{a'}[\alpha'(t)]\big\},
\ee
which depends only on the fixed ending points.
Substituting \Eqs{F2_debye} and (\ref{dotcalQ_debye}) into
\Eq{dotF1_debye} leads to
\be\label{dotF1_debyefinal}
  \partial_t{\cal F}_{\bm a} = {\cal C}_{\bm a}{\cal F}
 - \gamma_{\bm a}{\cal F}_{\bm a}
  - i\sum_{\bm b}{\cal Q}_b {\cal F}_{{\bm a}{\bm b}}   .
\ee

 We are now in the position to complete the
 IGF--COPI approach to the hierachical EOM
 for the multi-mode Drude-Debye dissipation of \Eq{debyeC}.
 First of all, we notice that due to the
 mathematical group nature implied
 in \Eq{dotF_debye} and \Eq{dotcalQ_debye},
 $(-i\ti{\cal Q}_{\bm a})$ constitutes
 the single influence generating functional for
 each pair of the Drude-Debye modes.
 The involving AIFs
 can be generically expressed
 as [cf.\ \Eqs{F1_debye} and (\ref{F2_debye})]
  \be\label{Fbfn}
  {\cal F}_{\sf n}\equiv
   \Big\{\prod_{\bm a}
  \big(-i\ti{\cal Q}_{\bm a}\big)^{n_{\bm a}} \Big\} {\cal F}.
  \ee
 Here ${\sf n}=(n_{\bm a}, n_{\bm b},\cdots)$ consists of
 a set of nonnegative integers.
  Denote also the index-set,
 ${\sf n}^{\pm}_{\bm a}\equiv (n_{\bm a}\pm 1,n_{\bm b},\cdots)$,
 that deviates from ${\sf n}$ only by
 changing the specified $n_{\bm a}$ to $n_{\bm a}\pm 1$.
 Note that the total number of nonnegative integers in
 the index set ${\sf n}$ is the same as that of
 the non-zero system--bath coupling mode pairs.

 The time derivative of ${\cal F}_{\sf n}$ can be carried out by
 using \Eq{dotcalQ_debye} and the first identity
 of \Eq{dotF_debye}, resulting in
 \be \label{dotFn_debye}
   \partial_t{\cal F}_{\sf n}=
   -\bigl(\sum_{\bm a} n_{\bm a} \gamma_{\bm a}\bigr){\cal F}_{\sf n}
      + \sum_{\bm a} \bigl(
        n_{\bm a} {\cal C}_{\bm a}
     {\cal F}_{ {\sf n}_{\bm a}^- }
     -i  {\cal Q}_a {\cal F}_{{\sf n}_{\bm a}^+ } \bigr) .
 \ee
  Define now the auxiliary propagators by [cf.\ \Eq{calGPI}]
 \bea \label{calGPIn_def}
   {\cal U}_{\sf n}(\bfalp,t;\bfalp_0,t_0)
 \equiv \int_{\bfalp_0}^{\bfalp}   \!\!  {\cal D}{\bfalp} \,
     e^{iS[\alpha]} {\cal F}_{\sf n}[\bfalp] e^{-iS[\alpha']} ,
 \eea
 which also define the related auxiliary density operators,
\be \label{rhosfn_def}
  \rho_{\sf n}(t)\equiv {\cal U}_{\sf n}(t,t_0)\rho(t_0).
\ee
Equations (\ref{dotFn_debye}) can then be recast into
 \be\label{Debyefinal}
    \dot\rho_{\sf n}=-\bigl(i{\cal L}+\sum_{\bm a}
     n_{\bm a} \gamma_{\bm a}\bigr)\rho_{\sf n}
      + \sum_{\bm a} \bigl(
        n_{\bm a} {\cal C}_{\bm a}
     \rho_{ {\sf n}_{\bm a}^- }
     -i  {\cal Q}_a \rho_{ {\sf n}_{\bm a}^+ } \bigr).
  \ee
 The involving ${\cal Q}_a$ and ${\cal C}_{\bm a}$,
 which in \Eqs{dotFn_debye} were given
 by \Eq{calQa} and \Eq{calCPI_debye}, respectively,
 at the fixed ending points of $\bfalp(t)=\bfalp$
 in the path-integral representation,
 are now defined in the operator level
  by \Eq{calLQ} and
 \be\label{calC_debye}
   {\cal C}_{\bm a}\hat O = -i(\eta_{\bm a}Q_{a'}\hat O
  - \eta^{\ast}_{\bm a}\hat OQ_{a'}).
 \ee
 Each of the summations in the rhs
 of \Eqs{Debyefinal} runs over all non-zero
 coupling mode pairs $\bm a=(aa')$.
  The initial conditions to \Eqs{Debyefinal}
 and the methods of infinite hierarchy truncation
 will be discussed later together with the general dissipation
 systems; see comments after \Eqs{calABC}
 and in \Sec{thresidue}.

    Equations (\ref{Debyefinal}) generalize
 the previous work on the single-mode Drude-Debye
 dissipation \cite{Tan89101,Tan914131,Ish053131,Tan06082001,%
Sha045053,Yan04216,Xu05041103}.
 It is noticed that at the second or higher tier,
 which is of the fourth or higher order in the
 system--bath coupling, contributions from
 different dissipative modes are no longer
 just additive. Some interesting phenomena such as
 co-tunneling \cite{Thi05146806}
 and co-dissipation (e.g., cooperative $T_2$-decoherence
 and $T_1$-relaxation processes)
 will therefore be anticipated and investigated
 elsewhere.

\section{Hierarchical equations of motion: General
 non-Markovian dissipation
}
\label{theom}

\subsection{Non-Markovian bath via parametrization}
\label{theomA}
  It is evident now that the construction
of hierarchical EOM for a general dissipative system should involve
a proper parametrization scheme for $C_{aa'}(t)$.
 It is required by the IGF--COPI structure  that
 all involving $\partial_t \ti Q_{\bm a}$ terms
 be contained within the hierarchy;
cf.\ \Eqs{dottiQ_debye}--(\ref{dotcalQ_debye}) and comments there.
 On the other hand, the relations of \Eq{CtsymFDT}, or more precisely
the fluctuation-dissipation theorem (FDT) should also be observed.

  In this work, we adopt a FDT-preserved parametrization
 scheme, in which the bath correlation functions
 $C_{\bm a}(t)$ for a general system at any temperature
 are completely characterized by a set of real
 parameters; see
 details in \App{thapp_para}.
 In particular, the parameters,
 $\{\gamma^{\bm a}_{\D},
  \gamma^{\bm a}_{k},\omega^{\bm a}_k;\, k\!=\!0,\!\cdots\!,\!K\}$
that will explicitly enter the final EOM,
together with the Matsubara frequencies
$\{\check\gamma_m\!\equiv\! 2\pi m/\beta;\; m\!=\!1,\!\cdots\!,\!M\}$
are all positive, except $\omega^{\bm a}_0\!\equiv\!0$.
For the latter use, we denote also
$\omega^{\prime{\bm a}}_k
 \!\equiv\! \delta_{k0}\gamma^{\bm a}_0+\omega^{\bm a}_k$.

 The final  FDT-preserved bath correlation functions assume
the following form [cf.\ \Eq{Cpara_app} where $M\rightarrow\infty$\,].
\be \label{Cpara}
  C_{\bm a}(t\!>\!0) = \eta^{\bm a}_{\D}e^{-\gamma^{\bm a}_{\D}t}
    \!+\! \sum_{j=0}^{2K+1}\!\!\eta^{\bm a}_{j}\phi^{\bm a}_{j}(t)
     \!+\! \sum_{m=1}^{M}\!\check\eta^{\bm a}_m
     e^{-\check{\gamma}_m t}.
\ee
Here, $\check{\gamma}_{m}=2\pi m/\beta$, and
\bsube \label{phifunc}
\bea
 &\phi^{\bm a}_{2k}(t)
 \equiv \cos(\omega^{\bm a}_{k}t)\exp(-\gamma^{\bm a}_{k}t),
\\
 &\phi^{\bm a}_{2k+1}(t)
 \equiv
  [\delta_{k0} \gamma^{\bm a}_{0}t
  +\sin(\omega^{\bm a}_{k}t)]\exp(-\gamma^{\bm a}_{k}t).
\eea
\esube
The $\eta$-coefficients in \Eq{Cpara} are all complex in general
due to the FDT; except those
 $\{\check\eta^{\bm a}_m\}$
 arising from the Matsubara contribution
are real in a canonical bath ensemble;
see \Eqs{etaD}--(\ref{Matsu_coef}).

Note that $\phi^{\bm a}_0(t) = e^{-\gamma^{\bm a}_0t}$ and
$\phi^{\bm a}_1(t) = \gamma^{\bm a}_0te^{-\gamma^{\bm a}_0t}$,
as $\omega^{\bm a}_0\equiv 0$.
Also (noting $\omega^{\prime\bm a}_{k}
   \equiv\delta_{k0}\gamma^{\bm a}_0+\omega^{\bm a}_k$)
\bsube \label{dotphiall}
\bea
   &\partial_t \phi^{\bm a}_{2k}(t)
   = -\gamma^{\bm a}_{k}\phi^{\bm a}_{2k}(t)
     -\omega^{\bm a}_{k}\phi^{\bm a}_{2k+1}(t),
\\
  &\partial_t \phi^{\bm a}_{2k+1}(t)
   = \omega^{\prime\bm a}_{k}\phi^{\bm a}_{2k}(t)
      -\gamma^{\bm a}_{k}\phi^{\bm a}_{2k+1}(t).
\eea
\esube
The above closed relations will be used in the following
 IGF--COPI construction of hierarchical EOM.

 In principle, the parametrized $C_{\bm a}$ in \Eq{Cpara}
 can be exact for arbitrary dissipation,
 if $K$ and $M$  in \Eq{Cpara} are large enough.
 The hierarchical EOM to be
 developed in the rest of this section will also be
 exact; but its size grows in a power law.
 The exact evaluation of complex dissipation would rapidly
 become extremely tedious.
 We will come back to this issue
 on how to incorporate the residue correction, due to
 the small difference between the exact
 $C_{\bm a}(t)$ (including the zero temperature case)
  and the practically used ones,
 into the final theory in \Sec{thresidue}.

\subsection{Hierarchical construction:
   Influence generating functionals}
\label{theomB}

 Note that $\ti Q_{\bm a}$ [\Eq{tilQFI}]
 appears {\it additive} with respect to the individual
 components of $C_{\bm a}$.
 The dissipation functional ${\cal R}$ [\Eq{calR_PI}
 with \Eq{ticalQa}] will also be additive.
 Moreover, in comparison with the first two terms in \Eq{Cpara},
 the Matsubara term possesses the special properties
 for its pre-exponential factors $\check\eta^{\bm a}_m$ being real
 and its time constants $\check\gamma_m$ being
 dissipation-mode independent. As results,
 the Matsubara contributions to ${\cal R}$ [\Eq{calR_PI}]
 can have the summation over $a'$ performed
 in the following construction of the hierarchical EOM formalism.

  To proceed, let us denote
 [$\phi^{\bm a}_{\D}(t)\equiv \exp(-\gamma^{\bm a}_{\D}t)$]
\be \label{tiQaj}
  \ti Q^{\bm a}_{j}(t;\{\alpha\}) \equiv
  \int_{t_0}^t\!d\tau \phi^{\bm a}_{j}(t-\tau)Q_{a'}[\alpha(\tau)],
\ee
and
\be \label{checkQm}
 \check Q^a_m(t;\{\alpha\}) \equiv \sum_{a'} \check\eta^{\bm a}_m
   \int_{t_0}^t\!d\tau e^{-\check{\gamma}_m (t-\tau)}Q_{a'}[\alpha(\tau)].
\ee
 The corresponding composite $\ti{\cal Q}$-functionals
 that specify the dissipation functional
 [cf.\ \Eq{calR_PI}] are denoted as
 \bsube \label{XY}
 \bea
    &X^{\bm a}_k(t;\{\bfalp\})
\equiv
   -i\big(\eta^{\bm a}_{2k}\ti Q^{\bm a}_{2k}
  - \eta^{{\bm a}\,\ast}_{2k}\ti Q^{\prime\bm a}_{2k}\big),
 \label{XinPI} \\
  &Y^{\bm a}_k(t;\{\bfalp\})
 \equiv
    -i\big(\eta^{\bm a}_{2k+1}\ti Q^{\bm a}_{2k+1}
   -\eta^{{\bm a}\,\ast}_{2k+1} \ti Q^{\prime\bm a}_{2k+1}\big),
 \label{YinPI}
 \eea
 \be
  Z^{\bm a}_{\D}(t;\{\bfalp\}) \equiv
  -i\big(\eta^{\bm a}_{\D}\ti Q^{\bm a}_{\D}
  - \eta^{{\bm a}\,\ast}_{\D}\ti Q^{\prime\bm a}_{\D}\big),
 \ee
 and [noting that $\check\eta^{\bm a}_m$ is real; see \Eq{Matsu_coef}]
 \be\label{checkcalQPI}
   \check Z^a_m(t;\{\bfalp\}) \equiv
  -i\big(\check Q^a_m- \check Q^{\prime\,a}_m\big).
 \ee
 \esube
 Included in each of the equation is also
 the factor of $(-i)$ for the sake of bookkeeping;
 e.g., $Z^{\bm a}_{\D}$ amounts to the
 $-i{\ti{\cal Q}}_{\bm a}$ in \Sec{thdebye}.

  The dissipative functional ${\cal R}$, by which
 \bsube\label{dotF}
 \be \label{dotFa}
  \partial_t{\cal F} = -{\cal R}{\cal F},
 \ee
 reads now [cf.\ \Eq{FV_FPhi}]
 \be \label{calRall}
  {\cal R}= i\sum_{\bm a} {\cal Q}_a Z^{\bm a}_{\D}
     + i\sum_{\bm a,k} {\cal Q}_a(X^{\bm a}_k+Y^{\bm a}_k)
     +i\sum_{a,m}{\cal Q}_a \check Z^a_m.
 \ee
 \esube
   Apparently, all composite $\ti{\cal Q}$-functionals,
 \Eqs{XY}, are influence generating functionals;
 they are however not completed.

 The crucial step in the IGF--COPI construction of hierarchical EOM
 is the time derivatives on these composite $\ti{\cal Q}$-functionals
 (cf.\  \Sec{thdebye}).
 Unlike the $Z$-functionals for the Drude and Matsubara components,
 the time derivatives of the $X$- and $Y$-functionals are
 closed together with two additional
 non-composite $\ti{\cal Q}$-functionals [cf.\ \Eq{dotphiall}].
 \bsube \label{XYbar}
 \bea
  \bar X^{\bm a}_k(t;\{\bfalp\})
\!\!&\equiv&\!\!
  -i\big(\eta^{\bm a}_{2k}\ti Q^{\bm a}_{2k+1}
  - \eta^{{\bm a}\,\ast}_{2k}\ti Q^{\prime\bm a}_{2k+1}\big),
\\
  \bar Y^{\bm a}_k(t;\{\bfalp\})
\!\!&\equiv&\!\!
   -i\big(\eta^{\bm a}_{2k+1}\ti Q^{\bm a}_{2k}
   -\eta^{{\bm a}\,\ast}_{2k+1} \ti Q^{\prime\bm a}_{2k}\big).
\eea
\esube
 The time derivatives of all involving $\ti{\cal Q}$-functionals
are obtained as [cf.\ \Eqs{dotphiall}
  and (\ref{dotcalQ_debye})]
\bsube \label{dotXYZ}
\bea
  \partial_t X^{\bm a}_k
  \!\!&=&\!\!
  {\cal A}^{\bm a}_k -\gamma^{\bm a}_{k} X^{\bm a}_k
   - \omega^{\bm a}_{k}\bar X^{\bm a}_{k},
\label{dotX1} \\
  \partial_t {\bar X}^{\bm a}_{k}
 \!\!&=&\!\!
    \omega^{\prime\,\bm a}_{k}X^{\bm a}_{k}
   -\gamma^{\bm a}_{k}\bar X^{\bm a}_{k};
\label{dotX2}
\eea
\bea
 \partial_t {\bar Y}^{\bm a}_k
 \!\!&=&\!\!
 {\cal B}^{\bm a}_k -\gamma^{\bm a}_{k} {\bar Y}^{\bm a}_k
   - \omega^{\bm a}_{k}Y^{\bm a}_{k},
\label{dotY1} \\
  \partial_t Y^{\bm a}_{k}
 \!\!&=&\!\!
  \omega^{\prime\bm a}_{k}{\bar Y}^{\bm a}_{k}
   -\gamma^{\bm a}_{k}Y^{\bm a}_{k};
\label{dotY2}
\eea
and
\be \label{dotZ}
 \partial_t Z^{\bm a}_{\D}={\cal C}^{\bm a}_{\D} -
 \gamma^{\bm a}_{\D} Z^{\bm a}_{\D}, \ \ \
 \partial_t \check Z^a_m
  = \check {\cal C}^a_m - \check{\gamma}_m \check Z^a_m\,.
\ee
\esube
Here
\bsube\label{calABC_PI}
\bea
  {\cal A}^{\bm a}_k
\!\!&\equiv&\!\!
  -i\left(\eta^{\bm a}_{2k}Q_{a'}-\eta^{{\bm a}\,\ast}_{2k}Q'_{a'}\right),
\label{calA_PI}
\\
  {\cal B}^{\bm a}_k
\!\!&\equiv&\!\!
  -i\left(\eta^{\bm a}_{2k+1}Q_{a'}
  - \eta^{{\bm a}\,\ast}_{2k+1}Q'_{a'}\right),
\label{calB_PI}
\\
 {\cal C}^{\bm a}_{\D}
\!\!&\equiv&\!\!
 -i\left(\eta^{\bm a}_{\D}Q_{a'}
     -\eta^{{\bm a}\,\ast}_{\D}Q'_{a'}\right),
\label{calCD_PI}
\\
 \check{\cal C}^a_m
\!\!&\equiv&\!\!
  -i\sum_{a'}\check\eta^{\bm a}_m
 \left(Q_{a'} - Q'_{a'}\right).
\label{calCm_PI}
\eea
\esube
Note that \Eq{calCD_PI} is identical to \Eq{calCPI_debye}.
In writing \Eq{calCm_PI}, we have used the property that
$\check\eta^{\bm a}_m$ is real; see \Eq{Matsu_coef}.

  The above six $(X\!Y\!Z)$-functionals constitute now a complete
 set of IGF.  The general expression for the
 AIFs in the hierarchy
 are then obtained as [cf.\ \Eq{Fbfn} and its comment above]
\bea\label{calFbl}
   {\cal F}_{\ind}
\!\!&=&\!\!
  \Bigl\{\prod_{{\bm a},k}
  \left[{(X^{\bm a}_k)}^{n^{\bm a}_{2k}}
        {(Y^{\bm a}_k)}^{n^{\bm a}_{2k+1}}
     {({\bar X}^{\bm a}_k)}^{\bn^{\bm a}_{2k}}
     {({\bar Y}^{\bm a}_k)}^{\bn^{\bm a}_{2k+1}}
  \right]
\nl \!\!&\ &\, \times
  \prod_{{\bm a}} (Z^{\bm a}_{\D})^{n^{\bm a}_{\D}}
  \prod_{a,m} {(\check Z^a_m)}^{\check n^a_m}
 \Bigr\} {\cal F}.
 \eea
The index in ${\cal F}_{\ind}$ is specified by
the involving nonnegative integers,
 \be \label{indn}
  {\ind}\equiv
 \left(n^{\bm a}_j, \bn^{\bm a}_j, n^{\bm a}_{\D},\check n^a_m;
    \; {\mbox{\footnotesize{$j=0,\!\cdots\!,\!2K+1;
    \; m=1,\!\cdots\!,\!M$}}}
  \right).
 \ee
 Therefore, the total number of the nonnegative integers
 in the index-set $\ind$ is $4(K+1)p + p + Mq$,
 where $q$ denotes the number of dissipative modes,
 and $p\leq q^2$ the number of nonzero dissipative mode pairs.

\subsection{Hierarchical equations of motion}
\label{theomC}

  The IGF--COPI approach to the hierarchical EOM can now
 be completed by taking the time derivative
 on ${\cal F}_{\ind}$ [\Eq{calFbl} with \Eqs{dotF} and (\ref{dotXYZ})].
 The final results read in terms of the auxiliary density operators as
  [cf.\ \Eqs{calGPIn_def} and (\ref{rhosfn_def})]
\be \label{dotrhon}
 \dot\rho_{\ind}=-(i{\cal L} +{\mathit\gamma_{\ind}})\rho_{\ind}
  +\rhonswap +\rhondown + \rhonup.
\ee
 The $\gamma$-term in \Eqs{dotrhon} arises
from the damping-terms of \Eqs{dotXYZ}. The resulting damping
constant is given by
 \bea\label{Gamind}
   {\mathit\gamma}_{\ind}
  \!\!&\equiv&\!\!
  \sum_{\bm a,k}
  (n^{\bm a}_{2k}+\bn^{\bm a}_{2k}+n^{\bm a}_{2k+1}
 +\bn^{\bm a}_{2k+1})\gamma^{\bm a}_k
\nl \!\!& &\!\!
 + \sum_{\bm a} n^{\bm a}_{\D}\gamma^{\bm a}_{\D}
 +\sum_{a,m} {\check n}^a_m \check{\gamma}_m .
 \eea

  The second term in \Eq{dotrhon}
stems from the (off-diagonal) swap-terms of \Eqs{dotX1}--(\ref{dotY2}).
It reads
 \bea\label{rhonswap}
 \rhonswap
 \!\!&=&\!\!
 -\sum_{\bm a,k} \omega_{k}^{\bm a}
    \big(
      n^{\bm a}_{2k}    {\rho}_{{\ind}_{\bm a,2k}^{\rightarrow}}
      + \bn^{\bm a}_{2k+1} {\rho}_{{\ind}_{\bm a,2k+1}^{\leftarrow}}
    \big)
 \nl &&
    +\sum_{\bm a,k} \omega_k^{\prime\bm a}
    \big(
          \bn^{\bm a}_{2k}  {\rho}_{{\ind}_{\bm a,2k}^{\leftarrow}}
        + n^{\bm a}_{2k+1}  {\rho}_{{\ind}_{\bm a,2k+1}^{\rightarrow}}
    \big).
 \eea
 The index-set ${\ind}^{\rightarrow}_{\bm a,j}$
 differs from $\ind$ of \Eq{indn}
 only by
 $(n^{\bm a}_j,\bn^{\bm a}_j) \rightarrow (n^{\bm a}_j-1,\bn^{\bm a}_j+1)$,
while ${\ind}^{\leftarrow}_{\bm a,j}$ by
$(n^{\bm a}_j+1,\bn^{\bm a}_j-1) \leftarrow (n^{\bm a}_j, \bn^{\bm a}_j)$,
at the specified $(\bm a,j)$.

 The third term in \Eq{dotrhon}
stems from the $(\cal{A,B,C})$-terms of \Eqs{dotXYZ}, while the
last term is from \Eq{dotF}.
They are the hierarchy-down and hierarchy-up contributions, respectively,
and given by
\bea\label{rhondown}
  \rhondown
 \!\!&=&\!\!
 \sum_{\bm a,k}\big(
   n^{\bm a}_{2k} {\cal A}^{\bm a}_k
       {\rho}_{{\ind}_{\bm a,2k}^{-}}
  + \bn^{\bm a}_{2k+1} {\cal B}^{\bm a}_k
       {\rho}_{{\bar{\ind}}_{\bm a,2k+1}^{-}}
    \big)
 \nl \!\!&&\!\!
    +\sum_{\bm a} n^{\bm a}_{\D}{\cal C}^{\bm a}_{\D}
      {\rho}_{{\ind}_{\bm a,\D}^{-}}
    +\sum_{a,m} \check n^a_m \check{\cal C}^a_m
        {\rho}_{{\check{\ind}}_{a,m}^-},
\eea
 and (\mbox{\footnotesize $j=$ D,\,$0,1,\cdots,2K+1$})
\be\label{rhonup}
 \rhonup \equiv -{\cal R}\rho_{\ind}
 = -i \sum_{\bm a,j} {\cal Q}_a   {\rho}_{{\ind}^+_{\bm a,j} }
    -i  \sum_{a,m} {\cal Q}_a   {\rho}_{{\check{\ind}}^+_{a,m} } .
\ee
The index-set ${\ind}_{\bm a,j}^\pm$ ($\bar{\ind}_{\bm a,j}^{-}$ or
$\check{\ind}_m^\pm$) differs from ${\ind}$ only by
changing the specified
$n^{\bm a}_j$ ($\bn^{\bm a}_j$ or $\check n^a_m$)
to $n^{\bm a}_j \pm 1$ ($\bn^{\bm a}_j - 1$ or $\check n^a_m \pm 1$).
Note that the $({\bar{\ind}_{\bm a,j}^{+}})^{\rm th}$--auxiliary reduced
density operators are not generated from \Eq{rhonup},
since $\bar X_k$ and $\bar Y_k$ do not appear in the
 dissipation functional ${\cal R}$ [\Eq{dotF}];
they are rather generated from the EOM for the
$({{\ind}_{\bm a,j}^{+}})^{\rm th}$--auxiliary
reduced density operators via the involving
swap {\footnotesize\{${\leftrightarrows}$\}}--terms
there [cf.\ \Eq{rhonswap}].

 In \Eq{rhondown}, ${\cal A}^{\bm a}_k$, ${\cal B}^{\bm a}_k$,
 ${\cal C}^{\bm a}_{\D}$, and $\check{\cal C}^a_m$ denote
 the reduced Liouville-space operator counterparts
  of \Eqs{calABC_PI} [cf.\ the comments above \Eq{calC_debye}].
\bsube \label{calABC}
\bea
  {\cal A}^{\bm a}_k\hat O
 \!\!&=&\!\!
   -i\big(\eta^{\bm a}_{2k}Q_{a'}\hat O
   - \eta^{\bm a\,\ast}_{2k}\,\hat OQ_{a'}\big) ,
\\
  {\cal B}^{\bm a}_k\hat O
 \!\!&=&\!\!
   -i\big(\eta^{\bm a}_{2k+1}Q_{a'}\hat O -
   \eta^{\bm a\,\ast}_{2k+1}\,\hat OQ_{a'}\big) ,
\\
  {\cal C}^{\bm a}_{\D}\hat O
\!\!&=&\!\!
   -i\big(\eta^{\bm a}_{\D}Q_{a'}\hat O
   - \eta^{\bm a\ast}_{\D}\,\hat OQ_{a'}\big) ,
\\
 \check{\cal C}_m^a\hat O
 \!\!&=&\!\!
  -i\sum_{a'}\check \eta^{\bm a}_m[Q_{a'},\hat O].
\eea
\esube
The reduced Liouville-space operator ${\cal Q}_a$ involved
in \Eq{rhonup} was given by \Eq{calLQ}.

 The initial conditions to \Eqs{dotrhon} are
 $\rho_{{\sf n}}(t_0)=\rho(t_0)\delta_{\sf{n0}}$,
 as inferred from their definitions, and
 $\rho_{\sf 0}(t)=\rho(t)$ is the reduced density
 operator of primary interest.
  Note that when the initial time $t_0\rightarrow-\infty$,
 the established Hierarchical EOM formalism imposes no approximation.
 The initial conditions are however $\dot{\rho}_{\ind}(t_0)=0$,
 where $t_0$ can be any time before applying the
 external time-dependent fields. The pulse-field induced dynamics
 will then be evaluated via \Eq{dotrhon}.

  The hierarchical EOM, \Eqs{dotrhon}--(\ref{calABC}), are exact for
 a general dissipation system that involves the parametrized
 bath correlation functions of \Eqs{Cpara}.
  The residue effect due to the small difference
 between the exact and the parametrized $C_{\bm a}(t)$
 on the final formalism will be carried out
 together with the hierarchy truncation via the principle of residue
 correction in the coming section.

  It is noted that with a proper re-arrangement, \Eqs{dotrhon} can be recast
 in the standard tridiagonal coupling form; see \App{thapp_frac}.
 Included there is also a
 variation of the above theory, expressed
 in terms of the hierarchical Green's
 functions and their related memory kernels
 and continued fraction formalism.
\section{Truncation and the principle of residue correction}
\label{thresidue}

\subsection{The principle of residue correction}
\label{thresidueA}

  To complete the EOM formalism, the
 infinite hierarchy in \Eqs{dotrhon}--(\ref{calABC}) should be
 truncated at a certain level, say the
 $(N_{\rm trun})^{\rm th}$--tier.
  The simplest way is to set all
 $\{\rho_{\ind}\}$ of the higher tiers to be zero.
 The resulting $\rho_{\sf 0}(t)=\rho(t)$ of the primary
 interest will be exact up to
 the $(2N_{\rm trun})^{\rm th}$--order in
 the system--bath coupling.
  Other truncation schemes related to
  different ways to the partial account for the
  higher orders effect,
  have also been proposed \cite{Xu05041103,%
  Tan89101,Tan914131,Ish053131,Tan06082001}.

  Here, we present the principle of residue correction,
  which itself is formally exact.
  It is applied not just to the truncation,
  but also to a recovery of the residue effect,
  due to the difference between the exact
  and the parametrized $C_{\bm a}(t)$ of \Eq{Cpara},
  at all levels of hierarchy.
  The hierarchy truncation that  concerns only at the anchor level
  will be treated in the next subsection.

  The principle of residue correction related to
  the finite difference between the exact
 and the parametrized ones,
  \be\label{Cres}
    \delta C_{\bm a}(t)\equiv C^{\text{exa}}_{\bm a}(t)- C_{\bm a}(t),
  \ee
 arises from the observation that the dissipation functional
 is additive [cf.\ \Eq{calR_PI} with \Eqs{ticalQa} and (\ref{tilQFI})].
\be  \label{calRPI_residue}
  {\cal R}^{\rm ex}[t;\{\bfalp\}] =
  {\cal R}[t;\{\bfalp\}] + \delta{\cal R}[t;\{\bfalp\}].
\ee
Here,
\be \label{delta_calR}
  \delta{\cal R}[t;\{\bfalp\}] =
  \sum_{\bm a}{\cal Q}_a[\bfalp]\;
  \delta\ti{\cal Q}_{\bm a}(t;\{\bfalp\}),
\ee
with $\delta{\ti{\cal Q}}_{\bm a}$ the same as \Eq{ticalQa}, but
associating with $\delta C_{\bm a}(t)$; i.e.\ [cf.\ \Eq{tilQa}]
\be\label{delta_tilQa}
  \delta{\ti Q}_{\bm a}(t;\{\alpha\})
  = \int_{t_0}^{t}\!d\tau\,
     \delta C_{\bm a}(t-\tau)Q_{a'}[\alpha(\tau)].
\ee
  The resulting exact influence functional of primary interest reads now
${\cal F}^{\rm ex}[\bfalp] = {\cal F}_{\rm resi}[\bfalp]{\cal F}[\bfalp]$,
with $ \partial_t {\cal F}_{\text{resi}}
     = -\delta{\cal R}\, {\cal F}_{\text{resi}}$.
  The AIFs defined in \Eq{calFbl} for the
 construction of the hierarchical EOM
should now be replaced by
 \be \label{Fnexa}
  {\cal F}^{\text{exa}}_{\ind}
   \equiv {\cal F}_{\ind}{\cal F}_{\text{resi}}.
 \ee
Its time derivative reads
 \be\label{dotFnexa}
  \partial_t {\cal F}^{\text{exa}}_{\ind}
 = (\partial_t{\cal F}_{\ind}){\cal F}_{\text{resi}}
   -{\delta\cal R} {\cal F}^{\text{exa}}_{\ind}.
 \ee
  The EOM for the corresponding exact $\rho_{\ind}$
 is then obtained as [cf.\ \Eqs{dotrhon}--(\ref{calABC})]
 \be \label{dotrhoexa}
 \dot\rho_{\ind} = -
   [i{\cal L}+\delta{\cal R}(t)+\gamma_{\ind}]\rho_{\ind}
  + \rhonswap + \rhondown + \rhonup.
\ee
 The residue correction due to $\delta C_{\bm a}(t)$ [\Eq{Cres}]
 is thus global; the resulting $\delta{\cal R}(t)$
 [\Eq{delta_calR}] modifies the individual hierarchical EOM
 at all levels.

 Apparently, $\delta{\cal R}$ of \Eq{delta_calR}
 and ${\cal R}$ of \Eq{calR_PI} are of the same mathematical
 structure.
 If their common operator-level expression
 were known and implementable readily without approximation,
 \Eq{POP_rho} would be used directly, without invoking
 the hierarchical EOM at all.
 However, it is only possible for special cases,
 such as the pure-dephasing limit (i.e., the case of $[H, Q_{\bm a}]=0$)
 or the driven Brownian oscillator system \cite{Yan05187}.

   The key idea behind \Eq{dotrhoexa} for the
 general non-Markovian dissipation is as followings.
 The total $C^{\rm exa}_{\bm a}$ is partitioned into
 two parts. One is the parametrized $C_{\bm a}$
 that carries most of the non-Markovian coupling strength
 and is expressed in the form of \Eq{Cpara}.
 Another is the residue $\delta C_{\bm a}$ that is
 assumed in the weak interaction regime.
 The strong dissipation due to parametrized $C_{\bm a}$ is
 treated via the hierarchical EOM approach developed
 in \Sec{theom} without approximation.
 In principle, the residue $\delta C_{\bm a}$ can be zero
 if $K$ and $M$ for the parametrized $C_{\bm a}$
 in \Eq{Cpara} are large enough.
 However, the size of hierarchical EOM grows in a power
 law as $K$ and $M$ increase;
 the exact evaluation of complex dissipation
 would rapidly become extremely tedious if not impossible.
 Therefore, it is a practical trade-off
 to have a nonzero but weak
 $\delta C_{\bm a}$, as long as its induced
 global residue $\delta{\cal R}$
 can be {\it accurately} described with a certain
 perturbative or nonperturbative formulation at the {\it operator level}.

  Let us start with the simplest one, the Markovian-residue limit,
 in which \Eq{delta_tilQa} reduces to
  \be   \label{delta_tilQa_mar}
   \delta{\ti Q}_{\bm a}
 \approx
   \delta{\bar C}_{\bm a}(t) Q_{a'}[\alpha(t)],
 \ee
 with
 \be\label{delbarC}
   \delta{\bar C}_{\bm a}(t)
   =\int_{t_0}^{t}\!d\tau\,\delta C_{\bm a}(t-\tau).
 \ee
   Note that the system variable
 $Q_{a'}$ in \Eq{delta_tilQa_mar}
 is now represented at the {\it ending time} $t$ of
 the path integral at which
 $\alpha(t)=\alpha$ and $\alpha'(t)=\alpha'$
 are {\it fixed}.  As results, \Eq{delta_calR} for
 the Markovian-residue dissipation
 can be expressed in the operator level as
 \be\label{delcalRmar}
   \delta{\cal R}(t) \hat O = \sum_{\bm a}
     [Q_a,
     \delta{\bar C}_{\bm a}\!(t)Q_{a'}\hat O
    - [\delta{\bar C}_{\bm a}\!(t)]^{\ast}\hat O Q_{a'}].
 \ee
 When $t\rightarrow \infty$, the above equation
 reduces to the Matsubara residue or finite temperature
 correction proposed by Ishizaki and Tanimura \cite{Ish053131,Tan06082001}.
 Their hierarchical EOM is the single-mode
 Drude dissipation version of \Eq{dotrhoexa}.

  The principle of residue correction is right rooted
 at the fact that $\delta{\cal R}$ [\Eq{delta_calR}]
 is of the same mathematical structure as
 time-local dissipation functional ${\cal R}(t)$
 [\Eq{calR_PI}]. As results, various well-established
 approximation schemes
 can be exploited for the superoperator $\delta{\cal R}$,
 as it describes weak residue dissipation.
 The most celebrated scheme may be the second-order time-local
 expression \cite{Yan05187,Xu029196,Yan002068,Yan982721},
 \bsube \label{calR2all}
 \be \label{calR2}
   \delta{\cal R}(t) \hat O \approx \sum_{\bm a}
  \big[Q_a,
    \delta\ti Q^{(2)}_{\bm a}(t)\hat O
  - \hat O [\delta{\ti Q}_{\bm a}^{(2)}(t)]^{\dg}
  \,\big],
 \ee
 where
 \be \label{tiQ2nd}
    \delta\ti Q^{(2)}_{\bm a} =
   \int_{t_0}^t\!d\tau\, \delta C_{\bm a}(t-\tau)
   e^{-i{\cal L}(t-\tau)} Q_{a'}.
 \ee
 \esube
 The dissipation-free propagator (assuming
 also time-independent system Hamiltonian for simplicity)
 is used here to connect $Q_{a'}[\alpha(\tau)]$
 in \Eq{delta_tilQa} to its value at the fixed
 ending path point of $\alpha(t)=\alpha$.
 The above operator level expressions are thus obtained \cite{Xu05041103}.
 In fact, \Eqs{calR2all} constitute the unified
 Bloch--Redfield--Fokker--Planck equation \cite{Yan002068,Yan05187}.

    Nonperturbative approximation schemes
  can also be applied to the evaluation of
 the global residue-induced $\delta{\cal R}(t)$.
 These include the non-interacting blip approximation
 or its variations \cite{Leg871,Wei99,Tho0115},
 the self-energy augmented methods \cite{Cui06449},
 and the use of the exactly solvable
 driven Brownian oscillator model \cite{Yan05187}
 to mimic the anharmonic system of interest.

\subsection{Hierarchy truncation via the principle of residue correction}
\label{thresidueB}

   Consider now the hierarchy truncation and its related issues.
   Apparently, the collection of anchor $\{\rho_{\sf N}\}$ should
   be properly specified. The hierarchy tier-up $\rhoNup$
  associating with an anchor $\rho_{\sf N}$ contains
  at least one component that goes beyond the desired
  anchoring confinement, and thus, is subject to truncation.

     The principle of residue correction is applied here
  by recognizing that the tier-up components can be recast
 as [cf.\ \Eqs{calFbl}]
 \bsube \label{rhoNupXYZ}
 \bea
   &\rho_{{\sf N}^{+}_{{\bm a},2k}}\!\!+\rho_{{\sf N}^{+}_{{\bm a},2k+1}}
 \!\!\! =
   (X^{\bm a}_{k} + Y^{\bm a}_{k}) \rho_{\sf N};
\label{rhoNupXY}
\\ %
   &\rho_{{\sf N}^{+}_{{\bm a},\D}} \!\!
  = Z^{\bm a}_{\D}\rho_{\sf N};
\ \ \
  \rho_{{\sf N}^+_{a,m}} \!\!
 = \check Z^a_m \rho_{\sf N}.
\label{rhoNupZ}
 \eea
 \esube
  The $(2k)^{\rm th}$-- and $(2k+1)^{\rm th}$--components
  for each dissipative mode pair ${\bm a}$ are grouped together
 for the reasons that they arise from the same
 term of the parametrized spectral density
 function  and they carry the same strength; cf.\ \App{thapp_para}.
  Therefore, they shall be treated equally as the truncation
 is concerned.

   The anchoring indexes can now be specified  for the individual
  constituents of the interaction bath correlation functions
  $C_{\bm a}(t)$ of \Eq{Cpara} as
\[
   N^{\bm a}_{\D},\,  N^{\bm a}_{k},\, {\check N}^a_{m};
  \ \ \text{with\ } k=0,\cdots,K;  \ m=1,\cdots,M.
\]
 The closed set of hierarchically coupled EOM
 will then contain those $\rho_{\ind}$, whose
  individual index-set consists of the nonnegative integers
 that are confined within
\bsube\label{confine_ind}
\bea
   &n^{\bm a}_{2k}+n^{\bm a}_{2k+1}+
      \bar n^{\bm a}_{2k}+\bar n^{\bm a}_{2k+1} \leq N^{\bm a}_k,
\label{confine_indA} \\
  &n^{\bm a}_{\D} \leq N^{\bm a}_{\D}, \ \ \
   \check n^{a}_m \leq \check N^a_m.
\label{confine_indB}
\eea
\esube
 The anchor index-set ${\sf N}$ in $\rho_{\sf N}$
 can now be defined as those with
 at least one of the above upper limits being reached.
 The constraint of \Eq{confine_indA} is consistent
 with the way of grouping in \Eq{rhoNupXY};
 Once the upper limit of \Eq{confine_indA}
 is reached, the associated tier-up
 $\rho_{{\sf N}^{+}_{{\bm a},2k}}$ and
 $\rho_{{\sf N}^{+}_{{\bm a},2k+1}}$
 are both subject to the truncation.

 The truncation can now be made based on
 the formally exact relations of \Eqs{rhoNupXYZ}.
  The involving $X^{\bm a}_{k}, Y^{\bm a}_{k},
 Z^{\bm a}_{\D}$ and $\check Z^a_m$ there
 are the reduced Liouville-space operators,
 whose path-integral representation counterparts
 were given by \Eqs{XY}.
 Taking \Eq{XinPI} for $X^{\bm a}_{k}$ for an example, its
 operator-level form  reads in contact with \Eq{rhoNupXY}  as
 \be \label{rhoNtrun}
    \rho_{{\sf N}^{+}_{\bm a},2k} \equiv X^{\bm a}_{k}\rho_{\sf N}
 = -i\big[\eta^{\bm a}_{2k}\ti Q^{\bm a}_{2k}\!(t)\rho_{\sf N}
        -\eta_{2k}^{{\bm a}\,\ast}
        \rho_{\sf N}\,\ti Q^{{\bm a}\dagger}_{2k}\!(t)
     \big] .
 \ee
  If $\rho_{{\sf N}^{+}_{{\bm a},2k}}$
 goes beyond the closed hierarchy, an approximated expression of
 $\ti Q^{\bm a}_{2k}\!(t)$, such as
 [cf.\ \Eq{tiQ2nd} and the comments there]
 \be \label{Qtrun}
   \ti Q^{\bm a}_{2k}\!(t)
  \approx \int_{t_0}^t\!d\tau\,
   \phi^{\bm a}_{2k}(t-\tau) e^{-i{\cal L}(t-\tau)}Q_{a'},
 \ee
 is adopted {\it locally} in the rhs of \Eq{rhoNtrun} to make the truncation.
 The resulting $\rho_{{\sf N}^{+}_{{\bm a},2k}}$
 retains the same leading term as the exact one,
 being of $(2N+2)^{\rm th}$--order in the specified
 system--bath coupling strength.
 Thus, the truncation with sufficiently large $N$ induces practically no error
 as far as the dynamics of $\rho(t)=\rho_{\sf 0}(t)$
 of primary interest is concerned.
 The construction of closed hierarchical EOM with residue correction
 is now completed.

\subsection{Discussions and comments}
\label{thresidueC}
  We re-emphasize here that
 when the values of $K,M$ in \Eq{Cpara} and the
 truncation anchor indexes in \Eq{confine_ind}
 are set to be sufficiently large, the residue effects on
 the primarily interested $\rho$ are effectively zero.
 The global residue correction and the truncation
 scheme introduced at both the global and the local truncation levels
 in the previous two subsections
 are made for the purpose of
 efficient evaluation of the reduced dynamics
 of primary interest.
 For example, one may simply terminate the hierarchical
 EOM by setting the aforementioned
 beyond-the-hierarchy $\rho_{{\sf N}^{+}_{{\bm a},2k}}=0$.
 The resulting $\rho$ of primary interest
 will be exact up to the $(2N)^{\rm th}$ order in
 the specified system--bath coupling strength.
 The improved truncation as \Eq{rhoNtrun}
 will lead to $\rho$ exact up to the $(2N+2)^{\rm th}$ order,
 rather than the $(2N)^{\rm th}$ order.
 The above two truncation schemes, which represent
 two different resummations for partially incorporating
 the higher--order effects on
 the reduced dynamics of primary interest,
 shall be of no practical difference when the
 convergence is reached.

   Note that various commonly used
 forms of the Bloch-Redfield theory and Fokker-Planck
 equations can be considered as the
 globally weak (residue)
 dissipation without invoking the hierarchical EOM at all.
 They can also be recovered
 if the second-order truncation scheme of \Eq{Qtrun}
 is applied to the primary tier of $\sf N=0$.
 On the other hand, if all second-tier auxiliary
 reduced density operators are set to be zero,
 the present hierarchical EOM formalism reduces to
 the second-order memory-kernel
 quantum dissipation theory.

   The global residue dissipation
 $\delta{\cal R}(t)$ introduced in the final formalism
 [\Eq{dotrhoexa}] is also for the purpose of efficiency.
 Note the number of the $n^{\rm th}$--tier auxiliary
 reduced density operators
 $\{\rho_{\ind}\}$ is $\frac{(n+P-1)!}{n!(P-1)!}$,
 where $P=4(K+1)p + p + Mq$ is the number of
 the nonnegative integers in the index ${\sf n}$;
 see comments after \Eq{indn} and in \App{thapp_frac} before \Eq{rvecn}.
 The size of the hierarchical EOM increases in
 a power law as the values of $K$ and $M$ for
 the parametrized $C_{\bm a}$ of \Eq{Cpara}.
 The residue correction is introduced  to
 reduce the required values of $K$ and $M$,
 as long as the induced residue weak dissipation
 $\delta{\cal R}(t)$ can be accurately evaluated.
 In this sense, the final residue-corrected
 hierarchical EOM formalism remains practically
 exact.  It is worthy to point out that
 the final formalism is even capable of treating
 the zero-temperature dissipation. At $T=0$,
 the Matsubara frequencies are
 all vanished and the exact $C^{\rm exa}_{\bm a}(t)$
 is given by \Eq{FDT_0K}.
 However, one can set a certain finite low temperature
 for the parametrized $C_{\bm a}$,
 and evaluate residue $\delta C_{\bm a}$--corrected
 hierarchical EOM dynamics,
 followed by the convergence test
 with a lower value of the parametrization temperature.

\section{Summary}
\label{thsum}

  In summary, we have constructed
 the residue-corrected hierarchical
 EOM formalism [\Eq{dotrhoexa} with \Eqs{Gamind}--\ref{calABC})].
 The construction consists of two main steps.
 The first is the IGF-COPI approach to
 the EOM formalism (\Sec{theom}), for the
 parametrized interaction bath correlation functions
 preserving the fluctuation-dissipation theorem.
 This step is itself exact, as the FDT-preserved
 parametrization involved can in principle
 represent arbitrary interaction bath correlation functions.
  The second step is the residue correction,
 concerning about the practical applicability of
 the present theory to a broad range of systems.
 The principle of residue correction
 (\Sec{thresidue}) is itself exact, rooted from the formal relation between
 the dissipation functional and the time-local
 dissipation superoperator; see \Sec{thpathintC}.
 The application of this principle to
 construct the global ($\delta C$-induced) or the local
 (truncation-induced) residue-correction
 invokes inevitably a certain approximation scheme.
 However, it can be made in a sufficiently
 accurate manner, as far as the primarily interested $\rho(t)$ is concerned
 (cf.\ \Sec{thresidueC}).
 As results, the residue-corrected hierarchical EOM formalism
 could be practically used in the
 study of general quantum dissipation systems interacting
 with arbitrary bath, arbitrary time-dependent external
 fields, and at any temperature, including
 $T=0$; see the last remark stated in \Sec{thresidueC}.

 The hierarchical EOM formalism
 may be relatively tractable,
 in comparison with the direct evaluation of
 the path--integral formalism \cite{Mak952430}.
 The memory effect in the quasiadiabatic propagator path integral
 method is described in terms of
 the nonlocality of the influence functional \cite{Mak952430},
 while it in the present differential formalism is resolved via a set of
 linearly coupled {\it time-local} auxiliary operators.
 The exponential-like series expansion of
 the parametrized $C_{\bm a}(t)$ [\Eq{Cpara}]
 can be considered as the separation of
 the time scales, and the resulting
 $\rho_{\sf n}$ is associated with
 the decay constant $\gamma_{\sf n}$.
 The different truncation anchor indexes
 [\Eq{confine_ind}] can therefore be identified
 to reduce the required number of equations,
 which otherwise grows exponentially if all
 $C_{\bm a}$-composites are treated equally.
 The weak residue correction at the global
 level provides an additional freedom
 to improve the numerical efficiency
 of the present EOM theory.

  Numerous existing quantum dissipation theories
 can be recovered from the present formalism.
 These include the Tanimura's hierarchical EOM
 for single-mode Drude dissipation \cite{Ish053131,Tan06082001}
 and the unified Bloch-Redfield-Fokker-Planck formulation
 \cite{Yan002068,Yan05187};
 see comments after \Eqs{delcalRmar} and
 the second paragraph of \Sec{thresidueC}, respectively.
 The equivalent hierarchical and
 continued-fraction Green's functions and
  memory kernels expressions are also presented;
 see \App{thapp_frac}.
   The application of the continued-fraction
 Green's function formalism to the two-state
 electron transfer system, with
 a single-mode Drude-Debye dissipation at
 the high--temperature limit,
 has been carried
 out recently, resulting in an analytical expression
 for the nonperturbative rate process \cite{Han0611438,Han06685}.

  Multi-mode dissipation is physically important.
 In the weak (second-order) dissipation regime,
 the effects of different system--bath coupling
 modes are additive. This simple property
 is no longer true in the strong dissipation regime,
 and it has been shown that the strong
 dissipation should include also the cases of long
 memory system--bath interactions \cite{Xu05041103}. 
 Cooperative dissipation, similar like the
 co-tunneling in quantum transport \cite{Thi05146806},
 could be a common phenomenon in reality.
  The present work is carried out
 in the bosonic canonical bath case. The extension
 to the grand canonical bath ensemble cases,
 including both fermion and boson statistics,
 will be treated elsewhere \cite{Jin2beJCP}.

\begin{acknowledgments}
 Support from the NNSF of China
 (No.\ 50121202, No.\ 20403016 and No.\ 20533060),
 Ministry of Education of China (No.\ NCET-05-0546),
 and the RGC Hong Kong (No.\ 604006)
 is acknowledged.
\end{acknowledgments}

\appendix

\section{Parametrization of bath correlation functions}
\label{thapp_para}

 The fluctuation-dissipation theorem (FDT)
relates the correlation functions $C_{aa'}(t)$
and the spectral density functions $J_{aa'}(\omega)$ by
\be \label{FDT}
 C_{aa'}(t) = \frac{1}{\pi} \int_{-\infty}^{\infty}\!d\omega
     \frac{e^{-i \omega t}J_{aa'}(\omega)}
     {1-e^{-\beta\omega}}  .
\ee
The spectral density functions $J_{aa'}(\omega)$
in a canonical ensemble satisfy in general
the symmetry relations
\be\label{symJ}
 J_{aa'}(\omega)= - J_{a'a}(-\omega) = J_{a'a}^{\ast}(\omega).
\ee
The FDT leads also to $J_{aa'}(0)=0$,
the spectrum positivity $J_{aa}(\omega > 0) \geq 0$,
and the Schwarz-inequality
$J_{aa}(\omega)J_{bb}(\omega) \geq |J_{ab}(\omega)|^2$.

  We adopt the following form of the
extended Meier--Tannor spectral-density
parametrization scheme \cite{Mei993365,Yan05187},
in which (setting $\omega^{\bm a}_0\equiv 0$)
\be \label{Jwpara}
  J_{\bm a}(\omega)
 = \frac{\zeta^{\bm a}_{\D}\omega}{\omega^2+(\gamma^{{\bm a}}_{\D})^2}
   + \sum_{k=0}^{K}
     \frac{\zeta^{\bm a}_k\gamma^{\bm a}_k\omega +
   i\bar \zeta^{\bm a}_k\omega^2}
 {|\omega^2 - (\omega^{\bm a}_k+i\gamma^{\bm a}_k)^2|^2}.
\ee
All involving parameters are real;
 $\omega^{\bm a}_{k\neq 0}$,
$\gamma^{\bm a}_k$ and $\gamma^{\bm a}_{\D}$ positive as well.
The symmetry relations in \Eq{symJ} require also
(noting that $\bar\zeta_{k}^{aa}=0$)
\[
 (\gamma^{ba}_{\D},\zeta^{ba}_{\D},
  \omega_k^{ba},\gamma_k^{ba}, \zeta_k^{ba},\bar\zeta_{k}^{ba})
 = (\gamma^{ab}_{\D},\zeta^{ab}_{\D},
  \omega_k^{ab},\gamma_k^{ab},\zeta_k^{ab},-\bar\zeta_{k}^{ab}).
\]
   The FDT [\Eq{FDT}] leads to the interaction
bath correlation functions of
[cf.\ Eq.\ (B4) of Ref.\ \onlinecite{Yan05187}]
\be \label{Cpara_app}
  C_{\bm a}(t\!>\!0) = \eta^{\bm a}_{\D}e^{-\gamma^{\bm a}_{\D}t}
    \!+\! \sum_{j=0}^{2K+1}\!\!\eta^{\bm a}_{j}\phi^{\bm a}_{j}(t)
     \!+\! \sum_{m=1}^{\infty}\!\check\eta^{\bm a}_m
      e^{-\check{\gamma}_m t},
\ee
with (noting $\omega^{\bm a}_0\equiv 0$)
\bsube \label{phifunc_app}
\be \label{phi1_app}
 \phi^{\bm a}_{2k}(t)
 \equiv \cos(\omega^{\bm a}_{k}t)\exp(-\gamma^{\bm a}_{k}t) ,
\ee
\be \label{phi2_app}
  \phi^{\bm a}_{2k+1}(t)
 \equiv
  [\delta_{k0} \gamma^{\bm a}_{0}t
  +\sin(\omega^{\bm a}_{k}t)]\exp(-\gamma^{\bm a}_{k}t).
\ee
\esube

 The first term in the rhs of \Eq{Cpara_app}
arises from the pole of the Drude term in \Eq{Jwpara}.
The involving coefficient is given by
\be \label{etaD}
   \eta^{\bm a}_{\D}=\frac{\zeta^{\bm a}_{\D}}{2}
   \left[\,\text{ctan}(\beta\gamma^{\bm a}_{\D}/2) - i\,\right].
\ee

 The second term in the rhs of \Eq{Cpara_app}
arises from the poles of the second term in \Eq{Jwpara}.
The involving functions $\phi_j(t)$, given by \Eqs{phifunc},
are chosen to be real in this work, rather than
complex exponential functions used in Ref.\ \onlinecite{Yan05187}.
Noting that $\phi^{\bm a}_0(t) = e^{-\gamma^{\bm a}_0t}$
 and $\phi^{\bm a}_1(t) = \gamma^{\bm a}_0t\phi^{\bm a}_0(t)$,
as inferred from $\omega^{\bm a}_0\equiv 0$.
The involving coefficients $\{\eta^{\bm a}_j\}$ are complex.
Those with even indexes are (including $k=0$ at which $\omega^{\bm a}_0=0$),
\bsube \label{eta_app}
\bea \label{eta1_app}
 \eta^{\bm a}_{2k}
\!\!&=&\!\!
 \frac{
    (\zeta^{\bm a}_k+\bar\zeta^{\bm a}_k)
   \gamma^{\bm a}_k \sinh(\beta\omega^{\bm a}_k)
  - \bar\zeta^{\bm a}_k\omega^{\bm a}_k
    \sin(\beta\gamma^{\bm a}_k)
 }{
  4\omega^{\bm a}_k\gamma^{\bm a}_k
   [\cosh(\beta\omega^{\bm a}_k) - \cos(\beta\gamma^{\bm a}_k)]
 }
\nl \!\!&\ & \!\!
  + i \frac{\bar\zeta^{\bm a}_k}{4\gamma^{\bm a}_k};
\eea
and those with odd indexes are
\be \label{eta2_app}
  \eta^{\bm a}_{1}
 =
  \frac{(\zeta^{\bm a}_0 + \bar\zeta^{\bm a}_0)}{4\gamma^{\bm a}_0}
   \left[\,\text{ctan}(\beta\gamma^{\bm a}_0/2) -i\,\right],
\ee
and ($2k+1=3,5,\cdots$)
\bea
  \eta^{\bm a}_{2k+1}
\!\!&=&\!\!
 \frac{
   \bar\zeta^{\bm a}_k\omega^{\bm a}_k
     \sinh(\beta\omega^{\bm a}_k)
    +(\zeta^{\bm a}_k+\bar\zeta^{\bm a}_k)\gamma^{\bm a}_k
    \sin(\beta\gamma^{\bm a}_k)
 }{
  4\omega^{\bm a}_k\gamma^{\bm a}_k
   [\cosh(\beta\omega^{\bm a}_k) - \cos(\beta\gamma^{\bm a}_k)]
 }
\nl \!\!&\ &\!\!
  -i\frac{\zeta^{\bm a}_k+\bar\zeta^{\bm a}_k}{4\omega^{\bm a}_k}.
\eea
\esube

 The last term in the rhs of \Eq{Cpara_app}
 arises from the Matsubara frequencies,
 \be \label{varpim}
   \check{\gamma}_{m} = 2\pi m/\beta; \ \ \ m=1,2,\cdots,
 \ee
which constitute poles of the $(1-e^{-\beta z})^{-1}$ factor in \Eq{FDT}.
The involving coefficients are
\be \label{Matsu_coef}
 \check\eta^{\bm a}_m \equiv -i(2/\beta)J_{\bm a}(-i\check{\gamma}_m)
 = (\check\eta^{\bm a}_m)^{\ast}; \ \ \ m=1,2,\cdots
\ee
Note that in general, while $J_{\bm a}\equiv J_{aa'}(\omega)$
can be complex when $a\neq a'$,
its analytical continuation to $J_{\bm a}(i\omega)$
is pure imaginary in a canonical ensemble system,
as inferred from \Eq{symJ}.
As results, $\check\eta^{\bm a}_m$ of \Eq{Matsu_coef} are all real.

 In principle, \Eqs{Cpara_app}--(\ref{Matsu_coef}) can be
 exact at an arbitrary finite temperature,
 if $K \rightarrow \infty$.
 In the hierarchical construction presented in \Sec{theom},
 $K$ is finite and the Matsubara terms $m=1,\cdots,M$ is also
 finite. The residue correction due to the  small
 difference between the exact $C_{\bm a}(t)$ and the
 parametrized ones with finite terms is
 considered in \Sec{thresidue}. By doing that,
 we can even treat the dissipation at $T=0$, where the
 FDT of \Eq{FDT} assumes
 \be \label{FDT_0K}
   C_{\bm a}(t) = \frac{1}{\pi} \int_{0}^{\infty}\!d\omega
     e^{-i \omega t}J_{\bm a}(\omega);
  \ \ \ \text{for\ \ }  T=0,
 \ee
 while the parametrized ones used
 in \Sec{theom} are at a certain
 finite low temperature.

\section{Recursive Green's functions, memory kernels,
  and continued fraction formalism}
\label{thapp_frac}

\subsection{The EOM in tridiagonal coupling matrix form}
\label{thapp_frac0}

 For the case of a single-mode Drude-Debye
 dissipation at high temperature, the EOM
 assumes the standard tridiagonal coupling matrix form;
 i.e., \Eqs{Debyefinal} in the special case of single mode.
 This property has been used to construct
 the continued fraction formalism \cite{Tan89101,Tan914131,Ish053131,Tan06082001}.
 The involving Green's functions are also identified,
 together with an interesting application to the establishment
 of an analytical expression for electron transfer rate
 processes \cite{Han0611438}.

   Here, we would like to extend our previous
 continued fraction--Green's function formalism
 to the present complex dissipation systems.
 To that end, we shall first recast \Eqs{dotrhon} in the
 standard tridiagonal coupling matrix form.
 This is done by considering the
 order of $\rho_{\sf n}$ depending on the hierarchy
 generating functionals.
 \[
   n_{\ind}\equiv
    \sum_{\bm a,k}(n^{\bm a}_{2k}+n^{\bm a}_{2k+1}
     +\bn^{\bm a}_{2k}+\bn^{\bm a}_{2k+1})
    +\sum_{\bm a} n^{\bm a}_{\D}
    +\sum_{a,m}{\check n}^a_m.
 \]
  Collect now all those $n^{\rm th}$--tier auxiliary density
 operators; i.e., $\rho_{\ind}$ that have $n_{\ind}=n$,
 and arrange them into a vector
 $\rvec_n\equiv\{{\rho}_{\ind};n_{\ind}=n\}$.
  The size of this vector is
 $\frac{(n+P-1)!}{n!\,(P-1)!}$,
 if there are no truncations involved;
 see \Sec{thresidueB}. Here $P$ is
 the number of nonnegative integers in the index-set $\ind$;
 see the comments after \Eq{indn}.

   The hierarchical EOM [\Eqs{dotrhon}]
 can then be re-arranged in the standard
 tridiagonal matrix form as
 \be\label{rvecn}
   \partial_t\rvec_n=-\bfLam_n\rvec_n
  -i\bfA_n\rvec_{n-1}-i\bfB_n\rvec_{n+1}.
  \ee
 Note that the swap ones $\rhonswap$ are now a part of
 $\rvec_n$, and the $\bfLam_n$ matrix elements
 are all numbers, except the Liouvillian ${\cal L}$, which
 can also be time dependent in the presence of
 external pulsed fields for example.
 All $\rhondown$ are arranged into $\rvec_{n-1}$
 with the Liouville-space operators
 defined in \Eqs{calABC} involved in the elements of $\bfA_n$.
 Those $\rhonup$ are now in $\rvec_{n+1}$ and
 the elements in $\bfB_n$ are according to \Eq{rhonup}.

   Note that $\rho_0(t) = \rho(t)$ is the reduced
 density operator of primary interest, and  $\bfLam_0=i{\cal L}$.
 The collection of the $n^{\rm th}$--tier auxiliary
 density operators have the leading contributions
 of the $(2n)^{\rm th}$--order in
 the overall system--bath couplings
 to the $\rho(t)$ of primary interest.
 The initial conditions to \Eqs{rvecn}
 are $\rvec_n(t_0) = \delta_{n0}\rho(t_0)$, as inferred
 from the definition.

\subsection{Green's functions versus  memory kernels}
\label{thapp_frac1}

  In terms of the propagators, by which
 \be \label{eq_appC2}
    \rvec_n(t)\equiv \Uvec_n(t,t_0)\rho(t_0),
 \ee
 \Eqs{rvecn} read [noting that $\Uvec_n(t_0,t_0)\equiv \delta_{n0}$]
\be\label{Uvecn}
   \partial_t\Uvec_n = -\bfLam_n\Uvec_n -i\bfA_n\Uvec_{n-1}
      -i\bfB_n\Uvec_{n+1}.
 \ee

Introduce now the hierarchical set of Green's functions in the reduced
 system Liouville space as follows.
 \bsube \label{eq_appC3}
 \bea
   \Uvec_0(t,t_0) \!\!&\equiv&\!\! \Gvec_0(t,t_0),
\label{eq_appC3a}\\
   \Uvec_n(t,t_0)
 \!\!&\equiv&\!\!
  -i\int^t_{t_0}\!\!d\tau \Gvec_n(t,\tau)\bfA_n\Uvec_{n-1}(\tau,t_0),
  \  \  n \geq 1.   \nl
\label{eq_appC3b}
 \eea
 \esube
 The initial conditions are $\Gvec_n(t_0,t_0)=1$.
 Substituting \Eqs{eq_appC3} into \Eq{Uvecn} leads then to
 \bsube \label{eq_appC5}
 \be\label{parGa}
  \partial_t \Gvec_n(t,t_0)
 = -\bfLam_n \Gvec_n(t,t_0) -\int^t_{t_0}\!\!d\tau
     \bfPi_n(t,\tau)\Gvec_n(\tau,t_0),
 \ee
 with the involving memory kernel of
 \be
  \bfPi_n(t,\tau) = \bfB_n\Gvec_{n+1}(t,\tau)\bfA_{n+1}.
 \ee
 \esube

 In particular, $\bfPi_0(t,\tau)\equiv\Pi(t,\tau)$ is
 the primary memory kernel, by which (noting that $\bfLam_0=i{\cal L}$)
 \be \label{cop}
   \dot\rho(t) = -i{\cal L}(t)\rho(t)
   - \int_{t_0}^t \! d\tau\, \Pi(t,\tau)\rho(\tau).
 \ee

  For the time-independent system Hamiltonian,
 $\Gvec_n(t,\tau)=\Gvec_n(t-\tau)$ and $\bfPi_n(t,\tau)=\bfPi_n(t-\tau)$.
  The above Green's functions (or memory kernels)
 can be resolved in the Laplace frequency domain
 via the continued fraction expression of \cite{Han0611438,Han06685}
 \be\label{confrac2}
  \hat\Gvec_n(s)
  =\frac{1}{s+\bfLam_n+
    \bfB_n\hat\Gvec_{n+1}(s)\bfA_{n+1}}\,.
 \ee
 Unlike the auxiliary propagators $\Uvec_n$ that couple
 with both $\Uvec_{n+1}$ and $\Uvec_{n-1}$,
  the Green's functions $\Gvec_n$ couple only with $\Gvec_{n+1}$.
 Thus, the evaluation of the reduced dynamics
  of primary interest could be more efficiently
  carried out in terms of the auxiliary Green's functions,
 especially when there is no time-dependent external fields.
 The continued fraction formalism, combined with
  the Dyson equation technique, has been
  recently applied to the simple electron transfer,
   a spin-boson system, in Drude-Debye
 solvents \cite{Han0611438,Han06685}.


\end{document}